\documentclass[10pt,showpacs,aps,pra,twocolumn]{revtex4-1}

\newcommand{\bear}{\begin{eqnarray}}
\newcommand{\eear}{\end{eqnarray}}
\newcommand{\be}{\begin{equation}}
\newcommand{\ee}{\end{equation}}
\newcommand{\beqn}{\begin{eqnarray}}
\newcommand{\eeqn}{\end{eqnarray}}
\newcommand{\beqnn}{\begin{eqnarray*}}
\newcommand{\eeqnn}{\end{eqnarray*}}

\begin{document}

\title{Variance uncertainty relations without covariances for three and four observables}

\author{V. V. Dodonov}
\email{vdodonov@fis.unb.br}
\affiliation{Institute of Physics and International Center for Physics, University of Brasilia,
P.O. Box 04455, Brasilia 70919-970, Federal District, Brazil
}

\begin{abstract}
New sum and product uncertainty relations, containing variances of three or four observables, but not
containing explicitly their covariances, are derived.
One of consequences is the new inequality, giving a nonzero lower bound for the product of two variances
in the case of zero mean value of the commutator between the related operators.
Moreover, explicit examples show that in some cases this new bound can be
better than the known Robertson--Schr\"odinger one.
\end{abstract}

\maketitle

\section{Introduction}

It is impressive that 90 years after the birth of the concept of uncertainty
relations in Quantum Mechanics \cite{Heis,Kennard}, this subject is still ``alive'',
in the sense that one can observe a burst of publications in this area,
devoted to generalizations of the traditional product inequalities for two observables
\cite{Robertson29,Robertson30}
\be
\sigma_{AA} \sigma_{BB} \ge  \frac14|
\langle [\hat{A},\hat{B}]\rangle|^2 
\label{unc3}
\ee
or its stronger version \cite{Robertson30,Schrod30}
\be
\sigma_{AA} \sigma_{BB} \ge \sigma_{AB}^2 + \frac14|
\langle [\hat{A},\hat{B}]\rangle|^2 \equiv G_{AB}^2.
\label{unc4}
\ee
Here
$\sigma_{AB} 
\equiv \frac12 \langle \{\delta\hat{A} , \delta\hat{B} \}\rangle$
and $\delta\hat{A} \equiv \hat{A} -\langle \hat{A}\rangle$.
Although the mainstreem of the current research is connected to the 
``entropic uncertainty relations'' (see \cite{Wehner10,BBR11,Coles17} for recent reviews),
several new inequalities containing {\em variances\/} of observables as measures of ``uncertainties''
have been discovered recently \cite{Rudnicki12,Mac-Pati,KW14,Yao15,Chen16}.
The goal of this article is to provide new families of relatively simple inequalities, containing on an equal footing
variances of three and four observables, but not containing explicitly any covariance 
(the specific choice of numbers $3$ and $4$ will become clear soon).
A remarkable consequence is the new inequality for the product of two variances, replacing inequality (\ref{unc3}) 
in the case of zero mean value of the commutator $[\hat{A},\hat{B}]$.

\section{Robertson's inequalities}

	The uncertainty relations for an arbitrary set of $N$ observables
were derived for the first time by Robertson \cite{Robertson34}.
Let us remind his scheme.
Consider $N$ arbitrary 
operators $\hat{z}_1$, $\hat{z}_2$, \ldots, $\hat{z}_N$, and construct the operator
$\hat{f} = \sum_{j=1}^N \alpha_j \delta\hat{z}_j $,
where $\alpha_j$ are arbitrary complex numbers.
The inequalities, which can be interpreted as generalized uncertainty relations, 
are the consequences of the fundamental 
inequality $\langle\hat{f}^{\dagger}\hat{f}\rangle\ge 0 $, that
must be satisfied for any pure or mixed quantum
state (the symbol $\hat{f}^{\dagger}$ means the Hermitian conjugated operator).
In the explicit form, this inequality is the condition of
positive semi-definiteness of the
quadratic form $\alpha^*_j F_{jm}\alpha_m $,
whose coefficients
$F_{jm} = \langle\delta\hat{z}_j^{\dagger} \delta\hat{z}_m\rangle$
form the Hermitian matrix $F =\Vert F_{jm}\Vert$.
One has only to use the known conditions of
the positive semi-definiteness of Hermitian quadratic forms 
to write down the explicit inequalities for the elements of 
matrix $F$. All such inequalities can be considered as
generalizations of inequality (\ref{unc4}) to the case of more than two operators. 
Many of them can be found, e.g., in review \cite{183} or Refs. \cite{Simon94,Sudar95,Trif02,Wun06,Ivan12}.
Applications to the problem of entanglement of continuous variable systems were
studied, e.g., in Refs. \cite{Serafini06,Nha07,Nha08,Lee14}.

If all operators $\hat{z}_j$ are Hermitian, then it is convenient to
split matrix $F$ as $F = X + iY$, where $X$ and $Y$ are
real symmetric and antisymmetric matrices, 
consisting of the elements
$
X_{mn}= \frac12 \left\langle\left\{\delta\hat{z}_m,
\delta\hat{z}_n\right\}\right\rangle $
and
$Y_{mn}= \frac1{2i} \left\langle\left[\hat{z}_m,\hat{z}_n\right]
\right\rangle $.
%
The symbols $\{,\}$ and $[\,,\,]$ mean the anticommutator and commutator, respectively. 
The fundamental inequality ensuring the positive semi-definiteness of matrix $F$ is
\be
 \det F =\det \Vert X + iY \Vert \ge 0.
\label{detF}
\ee
Unfortunately, this inequality is rather complicated for $N>2$ observables,
 because it contains, in addition to $N$ variances $X_{kk}$ 
and $N(N-1)/2$ mean values of commutators,
numerous sums and products of various combinations of $N(N-1)/2$ covariances $X_{jk}$ with $j\neq k$. 

For example, if $N=3$, then (\ref{detF}) can be written in the form
(see, e.g., \cite{Synge71,Qin16})
\beqn
&& X_{11}X_{22}X_{33} \ge   X_{11}\left(X_{23}^2 + Y_{23}^2\right)
+ X_{22}\left(X_{13}^2 + Y_{13}^2\right) 
\nonumber \\ &&
+ X_{33}\left(X_{12}^2 + Y_{12}^2\right) - 2X_{12}X_{23}X_{31}
\nonumber \\ &&
+ 2\left( X_{12}Y_{23}Y_{31} + X_{23}Y_{31}Y_{12} + X_{31}Y_{12}Y_{23} \right).
\label{unc17a} 
\eeqn
 In contradistinction to the
case of the Schr\"odinger inequality (\ref{unc4}), where removing the term $\sigma_{AB}$
from the right-hand size results in a simplified inequality (\ref{unc3}),
there is no possibility to simplify 
(\ref{unc17a}) by deleting all terms $X_{jk}$ with $j \neq k$,
since covariances $X_{jk}$ with $j \neq k$ can be positive or negative.
If such a simple trick could be done, then one would obtain the inequality
\be
X_{11}X_{22}X_{33} \ge X_{11}Y_{23}^2  + X_{22}Y_{13}^2 +X_{33}Y_{12}^2 .
\label{incor} 
\ee
But it is not satisfied, e.g., for the triple of dimensionless (scaled) operators 
(introduced in \cite{KW14}) $x$, $p$, and $\xi=x+p$, in the correlated coherent state \cite{DKM80}
\be
\psi_{\alpha}(x;\sigma,r) ={\cal N} \exp\left[-\,\frac{x^2}{4\sigma}
\left(1-\,\frac{ir}{\sqrt{1-r^2}}\right)
+\frac{\alpha x}{\sqrt{\sigma}} \right]
\label{psicorr}
\ee
with $\sigma=\hbar/\sqrt{3}$ and the correlation coefficient $r=-1/2$ (here ${\cal N}$ is the normalization factor). 
Indeed, we have in this case $Y_{jk}^2=\hbar^2/4$,
$\sigma_{xx}=\sigma_{pp}=\sigma_{\xi\xi}=\hbar/\sqrt{3}$, so that the left-hand side of
(\ref{incor}) equals $L=\hbar^3/(3\sqrt{3})$, while the right-hand side equals
$R=\hbar^3 \sqrt{3}/4=9L/4$.
Consequently, inequality (\ref{incor}) is wrong.

\section{Inequalities without explicit covariances for $N=3$}

The correct simplified inequality without covariances can be obtained using the scheme proposed
in \cite{Efimov76}.
The main idea is to extend the Hilbert space of states $|\psi\rangle$, considering the tensor products
$|\Psi\rangle =|\psi\rangle\otimes |\chi\rangle$, where $|\chi\rangle$ is an auxiliary spinor.
In this extended space we can introduce the operator
$ 
\hat{F} = \sum_{j=1}^3 \alpha_j \sigma_j \delta \hat{z}_j$,
where
 $\alpha_j$ are arbitrary {\em real\/} numbers
 and $\sigma_j$ are the standard $2\times 2$ Pauli matrices.
Then, using the anti-commutativity property of the Pauli matrices and performing averaging over the state $|\psi\rangle$, 
one can write
$
\langle\Psi| \hat{F}^{\dagger}\hat{F}|\Psi\rangle =
\langle\chi\ | {\cal A} |\chi\rangle
$
with the $2\times2$ Hermitian matrix (here $\sigma_0$ is the $2\times2$ unit matrix)
\beqn
{\cal A} &=& \left(\alpha_1^2 X_{11} + \alpha_2^2 X_{22} +\alpha_3^2 X_{33}\right)\sigma_0
\nonumber \\ &&
-2 \sigma_1 \alpha_2 \alpha_3 Y_{23} -2 \sigma_2 \alpha_3 \alpha_1 Y_{31} -2 \sigma_3 \alpha_1 \alpha_2 Y_{12}.
\label{calA}
\eeqn
 We see that matrix (\ref{calA})
 does not contain  covariances $X_{jk}$ with $j \neq k$.
Since $\langle\Psi| \hat{F}^{\dagger}\hat{F}|\Psi\rangle \ge 0$ for any physical state, matrix (\ref{calA})
must be positive semi-definite. 
Unfortunately, the analysis of this condition in Ref. \cite{Efimov76} suffered from some drawbacks, 
because the main result of that paper was the incorrect inequality (\ref{incor}).

The correct way is to use the condition $\det{\cal A} \ge 0$. It results in the inequality
\beqn
&& \alpha_1^2 X_{11} +\alpha_2^2 X_{22} + \alpha_3^2 X_{33} \nonumber \\ && \ge
2\left[\left(\alpha_1 \alpha_2 Y_{12}\right)^2 +\left(\alpha_2 \alpha_3 Y_{23}\right)^2 +\left(\alpha_1 \alpha_3 Y_{13}\right)^2
\right]^{1/2},
\label{gen}
\eeqn
which must hold for {\em arbitrary real numbers\/} $\alpha_1$, $\alpha_2$ and $\alpha_3$.
Looking for the most symmetric relations, let us choose $\alpha_1=\alpha_2=\alpha_3$.
Then we obtain the inequality
\be
 X_{11} + X_{22} +  X_{33} \ge
2\left[ Y_{12}^2 + Y_{23}^2 + Y_{13}^2
\right]^{1/2},
\label{sum3}
\ee
which is {\em stronger\/} than the consequence of the Robertson inequality (\ref{unc3})
\be
 X_{11} + X_{22} +  X_{33} \ge
 \left|Y_{12}\right| + \left|Y_{23}\right| + \left|Y_{13}\right|.
\label{sum3R}
\ee
In the special case of three canonical observables, $x$, $p$, and $\xi=x+p$, inequality (\ref{sum3})
was found in \cite{KW14}.

The choice $\alpha_k^2 =X_{kk}^n$ results in the inequality
\beqn
&& X_{11}^{n+1} + X_{22}^{n+1} +  X_{33}^{n+1} \nonumber \\ && \ge
2\left[ Y_{12}^2 X_{11}^n X_{22}^n + Y_{23}^2X_{33}^n X_{22}^n + Y_{13}^2X_{11}^n X_{33}^n
\right]^{1/2}.
\label{sum3n}
\eeqn
Of course, to use inequalities (\ref{sum3}) or (\ref{sum3n}) one should preliminary renormalize (rescale) 
observables $z_k$ in such a way that all of them acquire the same physical dimensions.

Wishing to find an inequality for the triple product $X_{11}  X_{22}   X_{33}$, let us choose
$\alpha_1^2 =X_{22}X_{33}$, $\alpha_2^2 =X_{11}X_{33}$ and $\alpha_3^2 =X_{22}X_{11}$.
Then the following correct inequality arises instead of (\ref{incor}):
\be
X_{11}  X_{22}   X_{33} \ge \frac49 \left(X_{11}Y_{23}^2  + X_{22}Y_{13}^2 +X_{33}Y_{12}^2\right).
\label{prod3}
\ee
It turns into the equality for the state (\ref{psicorr})
with $\sigma=\hbar/\sqrt{3}$ and  $r=-1/2$.

Applying the inequality $a+b \ge 2\sqrt{ab}$ to the right-hand side of (\ref{prod3}), we get the 
inequality $\xi^2 -2B\xi -\frac49 Y_{12}^2 \ge 0$, where $\xi \equiv \sqrt{X_{11}X_{22}}$
and $B\equiv 4|Y_{13}Y_{23}|/(9X_{33})$. Resolving this inequality with respect to the 
positive variable $\xi$, we arrive at the following inequality for the uncertainty
product $\Delta z_1 \Delta z_2 \equiv \sqrt{X_{11}X_{22}}$:
\be
\Delta z_1 \Delta z_2 \ge \sqrt{(2Y_{12}/3)^2 + B^2} +B.
\label{1122B}
\ee
This inequality is especially important if $Y_{12}=0$, when 
the standard Robertson uncertainty relation (\ref{unc3})
$\Delta z_1 \Delta z_2 \ge |Y_{12}|$ becomes useless. In this case a better inequality is
\be
\Delta z_1 \Delta z_2 \ge 8|Y_{13}Y_{23}|/(9X_{33}).
\label{Y120}
\ee

\subsection{Examples}
\label{sec-ex3}

A natural example of three observables is the set of three components $L_x$, $L_y$ and $L_z$
of the angular momentum vector ${\bf L}$. Inequality (\ref{sum3}) reads in this case as
\be
\langle {\bf L}^2\rangle - \langle {\bf L}\rangle^2 \ge \hbar\left|\langle {\bf L}\rangle\right|,
\label{L2L}
\ee
whereas inequality (\ref{prod3}) reads as
\be
L_{xx}L_{yy}L_{zz} \ge \hbar^2\left(L_{xx}L_x^2 + L_{yy}L_y^2 + L_{zz}L_z^2\right)/9,
\label{LxxLx2}
\ee
where $L_j \equiv \langle \hat{L}_j \rangle$ and 
$L_{jj} \equiv \langle \hat{L}_j^2 \rangle -\langle \hat{L}_j \rangle^2$.
If $L_z=0$, then inequality (\ref{Y120}) assumes the form
\be
\Delta L_x \Delta L_y \ge 2\hbar^2\left|L_x L_y/(9L_{zz}\right|.
\label{exLxLy}
\ee
Recently, many new uncertainty relations for the angular momentum operators were found
in \cite{Wun06,Rivas08, Damm15,Bjork16} (in addition to the set of inequalities collected in \cite{183}). 
But inequalities (\ref{L2L})-(\ref{exLxLy}) seem to be new.
To illustrate (\ref{exLxLy}), we have considered various superpositions of
 the  angular momentum $p$-states $|1,m\rangle$ with $m=1,0,-1$. 
It appears that the minimal ratio of the left-hand and
right-hand sides is achieved for the state
\be
|\psi\rangle = \frac12\left[|1,1\rangle + i|1,-1\rangle  +(1+i)|1,0\rangle\right],
\label{expsi}
\ee
possessing the following mean values and variances:
\[
L_x=L_y =\hbar/\sqrt{2}, \quad \Delta L_x = \Delta L_y = \hbar/2, \quad L_{zz}=\hbar^2/2.
\]
Then the right-hand side of (\ref{exLxLy}) equals $2\hbar^2/9$, and this value is only slightly smaller
than the value $\hbar^2/4$ of the left-hand side.
However, the Schr\"odinger inequality (\ref{unc4}) is more effective in this special case,
because $L_{xy}= -\hbar^2/4$ for the state (\ref{expsi}), so that $\Delta L_x \Delta L_y = |L_{xy}|$ exactly. 

Another interesting triple is
\be
\hat{z}_1 = \left(\delta\hat{p}\right)^2, \quad \hat{z}_2 = \left(\delta\hat{x}\right)^2, 
\quad \hat{z}_3 = \frac12\left(\delta\hat{p}\delta\hat{x}
+\delta\hat{x}\delta\hat{p}\right).
\label{triple}
\ee
Then 
\[
Y_{12}=-2\hbar\sigma_{px}, \quad
Y_{23}= \hbar\sigma_{xx}, \quad Y_{13}=-\hbar\sigma_{pp},
\] 
\[
X_{11} = \langle(\delta p)^4\rangle-\sigma_{pp}^2 \equiv \sigma_{4p}, \quad
X_{22} = \langle(\delta x)^4\rangle-\sigma_{xx}^2 \equiv \sigma_{4x}, 
\]
\[
X_{33} = \frac14\langle\left(\delta\hat{p}\delta\hat{x}
+\delta\hat{x}\delta\hat{p}\right)^2\rangle-\sigma_{px}^2.
\]
The simple Robertson inequality (\ref{unc3}) $\sigma_{4p}\sigma_{4x} \ge 4\hbar^2 \sigma_{px}^2$ 
becomes useless for states with
$\sigma_{px}=0$ (in particular, for any real wave function). In this case, it is better to use  
 inequality (\ref{Y120}), which assumes the form
\be
\sqrt{\sigma_{4p}\sigma_{4x}} \ge 32\hbar^2 \sigma_{pp}\sigma_{xx}
/[9\langle\left(\delta\hat{p}\delta\hat{x} +\delta\hat{x}\delta\hat{p}\right)^2\rangle].
\label{4p4x}
\ee
The right-hand side of (\ref{4p4x}) is obviously nonzero.
A simple illustration of the strength of inequality (\ref{4p4x}) can be done for arbitrary Gaussian states,
because all higher order statistical moments can be expressed in terms of (co)variances 
for this kind of states (see the Appendix for details). 
In particular,
\[
\sigma_{4p}=2\sigma_{pp}^2, \quad \sigma_{4x}=2\sigma_{xx}^2, 
\]
\[
\langle\left(\delta\hat{p}\delta\hat{x} +\delta\hat{x}\delta\hat{p}\right)^2\rangle = 
4\sigma_{pp}\sigma_{xx} + 8\left(\sigma_{xp}\right)^2 +\hbar^2.
\]
Consequently, the left-hand side of (\ref{4p4x}) equals $2\sigma_{pp}\sigma_{xx}$.
On the other hand, if $\sigma_{px}=0$, then the right-hand side equals 
${32\hbar^2 \sigma_{pp}\sigma_{xx}}/\left[{9\left(4\sigma_{pp}\sigma_{xx} +\hbar^2\right)}\right]$. 
For pure uncorrelated quantum Gaussian states we have $\sigma_{pp}\sigma_{xx}=\hbar^2/4$,
so that the ratio of the left-hand side of inequality (\ref{4p4x}) to its right-hand side
equals $9/8$. 
Moreover, inequality (\ref{4p4x}) turns out better than the Schr\"odinger inequality (\ref{unc4})
in this special case. Indeed,  for any Gaussian state one has (see the Appendix)
\[
X_{12} \equiv  \frac12 \langle (\delta\hat{p})^2(\delta\hat{x})^2 
+(\delta\hat{x})^2(\delta\hat{p})^2\rangle = \sigma_{pp}\sigma_{xx} + 2\sigma_{px}^2 -\hbar^2/2,
\]
so that  the value $\sqrt{X_{11}X_{22}}= \hbar^2/2$
is {\em twice\/} bigger than $|X_{12}| =\hbar^2/4$ for all pure Gaussian states with $\sigma_{px}=0$.
Note that $X_{12} <0$ in this case, due to the non-commutativity of the coordinate and
momentum operators.

\section{Four observables}

The scheme used in the preceding section can be generalized to sets of four arbitrary Hermitian operators,
if one replaces three $2\times2$ Pauli's matrices $\sigma_k$ with four Hermitian $4\times4$ Dirac's matrices
\[
\gamma_k =\left\Vert
\begin{array}{cc}
0 & \sigma_k
\\
\sigma_k & 0
\end{array}
\right\Vert, 
\quad k=1,2,3, 
\qquad
\gamma_4 =\left\Vert
\begin{array}{cc}
I_2 & 0
\\
0 & -I_2
\end{array}
\right\Vert, 
\]
where $I_n$ is the $n\times n$ unit matrix.
Then 
\be
 \gamma_m \gamma_n + \gamma_n\gamma_m = 2 I_4 \delta_{mn},  \quad
m,n = 1,2,3,4,
\label{antigam}
\ee
\[
\gamma_j \gamma_k - \gamma_k\gamma_j = 2i \epsilon_{jkl}
\left\Vert
\begin{array}{cc}
\sigma_l & 0
\\
0 & \sigma_l
\end{array}
\right\Vert,
\quad j,k,l=1,2,3,
\]
\[
\gamma_k \gamma_4 - \gamma_4\gamma_k = 2
\left\Vert
\begin{array}{cc}
0 & -\sigma_k
\\
\sigma_k & 0
\end{array}
\right\Vert, 
\quad k=1,2,3.
\]
Here $\epsilon_{jkl}$ is totally antisymmetric tensor with $\epsilon_{123}=1$.

Let us consider the operator $\hat{f} = \sum_{k=1}^4 \xi_k \hat{z}_k \gamma_k$, where 
$\xi_k$ are arbitrary {\em real\/} coefficients and $ \hat{z}_k$ arbitrary Hermitian operators.
Then  the condition $\langle \hat{f}^{\dagger}\hat{f}\rangle \ge 0$ can be written as the condition of
positive semi-definiteness of the Hermitian $4\times4$ block matrix
\be
F =\left\Vert
\begin{array}{cc}
A & B^{\dagger}
\\
B & A
\end{array}
\right\Vert,
\label{FAB}
\ee
\[
A  
= \left\Vert
\begin{array}{cc}
g_{\xi} -2\xi_1 \xi_2 Y_{12} & -2\xi_2 \xi_3 Y_{23} + 2i\xi_3 \xi_1 Y_{31}
\\
-2\xi_2 \xi_3 Y_{23} - 2i\xi_3 \xi_1 Y_{31} & g_{\xi} +2\xi_1 \xi_2 Y_{12}
\end{array}
\right\Vert, 
\]
\[
B 
= 2\left\Vert \begin{array}{cc}
 i\xi_3 \xi_4 Y_{34} & i\xi_1 \xi_4 Y_{14} +\xi_2 \xi_4 Y_{24}
\\
i\xi_1 \xi_4 Y_{14} -\xi_2 \xi_4 Y_{24} & -i\xi_3 \xi_4 Y_{34}
\end{array}
\right\Vert,
\]
\[
g_{\xi} = \sum_{k=1}^4 \xi_k^2 X_{kk}.
\]
The covariances $X_{jk}$ with $j \neq k$ do not appear due to the anti-commutation relations
(\ref{antigam}). The positivity condition containing all variances $X_{kk}$ and mean values of
commutators $Y_{jk}$ is $\det F \ge 0$. After some algebra, it can be written in the following compact
form:
\be
\det F = \left(g_{\xi}^2 - 4V_{\xi}\right)^2 - 64\left(\xi_1 \xi_2 \xi_3 \xi_4\right)^2 \Lambda^2 \ge 0,
\label{main-4}
\ee
\[
V_{\xi} = \sum_{j<k} \left(\xi_j \xi_k Y_{jk}\right)^2, \quad
\Lambda =  \left|Y_{12}Y_{34} + Y_{23} Y_{14} + Y_{31}Y_{24}\right|.
\]
Note that  $\Lambda$ is invariant with respect to the ordering of indexes, due to the
property $Y_{jk}=-Y_{kj}$.

Taking all $\xi_k=1$ (this means that the dimensions of operators $\hat{z}_k$ should be made equal by
means of some scaling transformations), we get the inequality
\be
 \left|\left(\sum_{k=1}^4 X_{kk}\right)^2 - 4\sum_{j<k} Y_{jk}^2\right| 
\ge 8\Lambda. 
\label{main-xi1}
\ee

Choosing 
$
\xi_j^2 = X_{kk} X_{mm} X_{nn}$ with $j \neq k \neq m \neq n$,
one can transform (\ref{main-4}) to the form
\be
\left(4 P-\Psi\right)^2 \ge 4P\Lambda^2, \qquad
P=X_{11}X_{22} X_{33} X_{44},
\label{4P}
\ee
\beqnn
\Psi &=& Y_{12}^2 X_{33}X_{44} +Y_{13}^2 X_{22}X_{44} +Y_{14}^2 X_{33}X_{22}
\\ &&
+Y_{23}^2 X_{11}X_{44} +Y_{24}^2 X_{33}X_{11} +Y_{34}^2 X_{11}X_{22}.
\eeqnn
Resolving inequality (\ref{4P}) with respect to variable $P$, we arrive at the new inequality
\be
8X_{11}X_{22} X_{33} X_{44} \ge 2\Psi +\Lambda^2 +|\Lambda|\sqrt{4\Psi +\Lambda^2} 
\label{Pgr}
\ee
(the second solution, giving an upper bound for $P$, is unphysical).
One can check that the equality sign in (\ref{Pgr}) is achieved, e.g., for the 
coordinates and momenta in the minimum uncertainty states of two uncoupled harmonic oscillators. 
Note that 
\be
\Psi \ge \Psi_* =2\left(Y_{12}^2 Y_{34}^2 + Y_{23}^2 Y_{14}^2 + Y_{31}^2 Y_{24}^2\right)
\label{Psi*}
\ee
as a consequence of the standard uncertainty relation (\ref{unc4}).
Therefore one can get rid of variances in the right-hand side of (\ref{Pgr}), replacing $\Psi$ with $\Psi_*$.

At this point it is worth comparing (\ref{Pgr}) with one of Robertson's inequalities \cite{Robertson34} 
\be
X_{11}X_{22}\ldots X_{NN} \ge  \det X \ge \det Y.
\label{unc10}
\ee
It is useless for $N=3$ (because $\det Y \equiv 0$ for any antisymmetric $3\times3$ matrix $Y$).
But for $N=4$ we have
\beqn
\det(Y) &=& Y_{12}^2 Y_{34}^2 + Y_{23}^2 Y_{14}^2 + Y_{31}^2 Y_{24}^2
-2Y_{12}Y_{13}Y_{24}Y_{34}
\nonumber \\ &&
-2 Y_{12}Y_{14}Y_{23}Y_{43} -2Y_{13}Y_{14}Y_{32}Y_{42} \equiv \Lambda^2.
\label{detY4}
\eeqn
For the sets of coordinates and momenta operators, the right-hand side of (\ref{unc10})
coincides with the right-hand side of (\ref{Pgr}), {\em provided\/} $\Psi$ is replaced by $\Psi_*$.
But if such a replacement is not done, then inequality (\ref{Pgr}) is {\em stronger\/}
than  (\ref{unc10}).

For example, let us consider the set $z_1 =x$, $z_2=p_x$, $z_3=y$, $z_4=p_y$, and the 
pure quantum state
\be
\psi(x,y)= {\cal N}\exp\left( -\frac{a}{2} x^2 - b xy - \frac{c}{2} y^2\right),
\label{psib}
\ee
where ${\cal N}$ is the normalization factor, and all coefficients, $a$, $b$, and $c$, are
real numbers, satisfying the restrictions $a>0$, $c>0$, and
 $D \equiv ac - b^2 >0$. Then 
\[
X_{11} =  \frac{c}{2D}, \quad  X_{33}  = \frac{a}{2D}, \quad
X_{22}= \frac12 a\hbar^2, \quad X_{44} = \frac12 c\hbar^2,
\]
so that
\[
P = \frac{(ac)^2 \hbar^4}{16 D^2}, \qquad \Psi = \frac{ac \hbar^4}{8 D}, 
\qquad
\Lambda = \frac{\hbar^2}{4}.
\]
If $b=0$, then we have equalities in both relations, (\ref{Pgr}) and (\ref{unc10}).
But if $b \neq 0$, and especially if $D \ll ac$, then  (\ref{unc10}) becomes very weak.
On the other hand, inequality (\ref{Pgr}) shows that the product $P$ must be much bigger
than $\hbar^4/16$ in this case (although the right-hand side of (\ref{Pgr}) is much smaller
than the left-hand side).

\section{Conclusion}

The main results of this paper are new inequalities (\ref{gen}), (\ref{sum3}), and (\ref{prod3}) for arbitrary
sets of three Hermitian operators, and inequalities (\ref{main-4}), (\ref{main-xi1}), and (\ref{Pgr})
for arbitrary sets of four Hermitian operators. These inequalities give lower bounds for the sums and
products of three or four variances in terms of the mean values of commutators,
but they do not contain explicitly the covariances between the observables.
Therefore the new inequalities are simpler than the known Robertson's inequalities for several observables.
(Note, however, that covariances can enter the inequalities implicitly, through the mean values of commutators, 
as happened in the second example  of Sec. \ref{sec-ex3}.)
  The ``magic numbers'' 3 and 4 arise due to the existence
of three anticommuting $2\times2$ Pauli's matrices and four anticommuting $4\times4$ Dirac's matrices.

An important consequence of new relations is inequality (\ref{Y120}), which shows that even if the mean
value of the commutator between two operators is zero, nonetheless, the product of variances of
the corresponding observables must be nonzero, if these observables are parts of some extended
system.
Perhaps, it is worth mentioning that all inequalities derived in this article remain valid, if one
uses unshifted operators $\hat{z}_j$ instead of $\delta\hat{z}_j$ in the construction of operator $\hat{F}$.
Then simplified versions of the new inequalities can be written, using the definition 
$X_{jj} =\langle \hat{z}_j^2\rangle$.

\begin{acknowledgments}
A partial support of the Brazilian funding agency 
Conselho Nacional de Desenvolvimento Cient\'{\i}fico e Tecnol\'ogico
(CNPq) is acknowledged.
\end{acknowledgments}

\appendix

\section{}

The variances of the triple (\ref{triple}) correspond to the fourth order moments
of coordinates and momenta. Such moments can be calculated easily for any Gaussian state, because its Wigner functions 
(in the single space dimension for simplicity)
$W\left(x,p_x\right)$ is also Gaussian,
so that one can use classical formulas 
for average values of the Gauss distributions
(with some modifications due to the non-commutativity of the coordinate 
and momentum operators). The details can be found, e.g., in \cite{183-3}. 

Consider four operators (not necessarily different) $\hat{A}$, $\hat{B}$, $\hat{C}$, and $\hat{D}$ (with zero mean values),
where each of them can be either $\delta\hat{x}$ or $\delta\hat{p}$.
Then the mean value of the {\em symmetrical\/} (or Wigner--Weyl) product of these operators is given
by the formula (see, e.g., \cite{Hill84}) 
\be
\langle{ABCD}\rangle_W = \int W\left(x,p\right)\,ABCD\,dxdp/(2\pi\hbar).
\label{meanW}
\ee
The meaning of symbol $\langle{ABCD}\rangle_W$ is the following:
this is the quantum mechanical mean value of the sum of all different products of operators $\hat{A},\hat{B},\hat{C},
and \hat{D}$, taken in all possible orders, divided by the number of terms. 
For example, if $\langle \hat{x}\rangle =\langle \hat{p}\rangle =0$, then
\[
\langle x^2 p^2 \rangle_W = \frac16 \langle \hat{x}^2 \hat{p}^2 +\hat{p}^2\hat{x}^2 +
\hat{x}\hat{p}\hat{x}\hat{p} + \hat{p}\hat{x}\hat{p}\hat{x} +\hat{x}\hat{p}^2\hat{x}
+\hat{p}\hat{x}^2\hat{p} \rangle.
\]
Mean values of concrete products of operators in predefined
orders can be expressed in terms of symmetrical mean values with the aid of commutation relations.
For example (if $\langle \hat{x}\rangle =\langle \hat{p}\rangle =0$),
\[
\langle x^2 p^2 \rangle_W = \frac12 \langle \hat{x}^2 \hat{p}^2 +\hat{p}^2\hat{x}^2 \rangle + \frac{\hbar^2}{2},
\]
\[
\langle \left(\hat{x} + \hat{p}\right)^2 \rangle = 2 \langle \hat{x}^2 \hat{p}^2 +\hat{p}^2\hat{x}^2 \rangle + 3\hbar^2.
\]

Since the Gaussian Wigner function is positive, one can consider it as a classical probability distribution and
apply  the classical formulas for the Gaussian probabilities to the right-hand side of (\ref{meanW}).
The final result is the following formula of decoupling the fourth-order moments into the sums of products
of the second-order moments:
\be
\langle{ABCD}\rangle_W =
\overline{AB}\cdot\overline{CD} + \overline{AC}\cdot\overline{BD} + \overline{AD}\cdot\overline{BC},
\label{basic}
\ee
where $\overline{AB} \equiv \frac12\langle \hat{A}\hat{B} +\hat{B}\hat{A} \rangle$ (remember that we suppose here
that $\langle \hat{A}\rangle =\langle \hat{B}\rangle =0$).
In particular, taking $\hat{A}=\hat{B}=\hat{C}=\hat{D}=\delta\hat{x}$ we arrive at the known formula
\be
 \langle (\delta\hat{x})^4 \rangle = 3 \left(\sigma_{xx}\right)^2.
\label{x4}
\ee
Another formula used in the main text is
\be
\langle (\delta x)^2 (\delta p)^2 \rangle_W = \sigma_{xx} \sigma_{pp} +2\left(\sigma_{xp}\right)^2.
\label{xp}
\ee

\end{document}